\documentclass{article}
\usepackage[total={6in,8.5in}]{geometry}

\usepackage[authoryear]{natbib}
\usepackage{hyperref}
\usepackage{amsmath}
\usepackage{amssymb}
\usepackage{lmodern}
\usepackage{graphicx}
\usepackage{authblk}

\usepackage[capitalize]{cleveref}
\creflabelformat{equation}{#2#1#3}

\title{Bayesian Inference of Transition Matrices from Incomplete Graph Data with a Topological Prior}

\author{
Vincenzo Perri
$^{1\star}$,
Luka V Petrovi\'c
$^{1\star}$,
Ingo Scholtes$^{2,1}$
\\
{\small
$1$ Data Analytics Group, Department of Informatics, University of Zurich, Switzerland\\
$2$ Chair of Machine Learning for Complex Networks, Center for Artificial Intelligence and Data Science, Julius-Maximilians-Universit\"at W\"urzburg, Germany\\
$\star$ equal contribution
}
}

\begin{document}
\maketitle

\begin{abstract}
Many network analysis and graph learning techniques are based on discrete- or continuous-time models of random walks.
To apply these methods, it is necessary to infer transition matrices that formalize the underlying stochastic process in an observed graph.
For weighted graphs, where weighted edges capture observations of repeated interactions between nodes, it is common to estimate the entries of such transition matrices based on the (relative) weights of edges.
However in real-world settings we are often confronted with incomplete data, which turns the construction of the transition matrix based on a weighted graph into an \emph{inference problem}.
Moreover, we often have access to additional information, which capture topological constraints of the system, i.e. which edges in a weighted graph are (theoretically) possible and which are not.
Examples include transportation networks, where we may have access to a small sample of passenger trajectories as well as the physical topology of connections, or a limited set of observed social interactions with additional information on the underlying social structure.
Combining these two different sources of information to reliably infer transition matrices from incomplete data on repeated interactions is an important open challenge, with severe implications for the reliability of downstream network analysis tasks.

Addressing this issue, we show that including knowledge on such topological constraints can considerably improve the inference of transition matrices, especially in situations where we only have a small number of observed interactions.
To this end, we derive an analytically tractable Bayesian method that uses repeated interactions and a topological prior to perform data-efficient inference of transition matrices.
We compare our approach against commonly used frequentist and Bayesian approaches both in synthetic data and in five real-world datasets, and we find that our method recovers the transition probabilities with higher accuracy.  
Furthermore, we demonstrate that the method is robust even in cases when the knowledge of the topological constraint is partial.
Lastly, we show that this higher accuracy improves the results for downstream network analysis tasks like cluster detection and node ranking, which highlights the practical relevance of our method for interdisciplinary data-driven analyses of networked systems.
\end{abstract}

\section{Introduction}
\label{sec:introduction}

Graph models of relational data have become a cornerstone in the analysis of complex systems~\citep{boccaletti2006complex} and an important foundation for the application of machine learning to graph-structured data from social, technical, and biological systems~\citep{bronstein2017geometric}.
Many network analysis and graph learning techniques are based on discrete- or continuous-time models of random walks \citep{masuda2017random} such as, e.g., community detection algorithms like InfoMap~\citep{rosvall2008maps} or WalkTrap \citep{pons2006computing}, node ranking techniques like PageRank~\citep{page1999pagerank}, neural graph embeddings like DeepWalk \citep{perozzi2014deepwalk} or node2vec~\citep{grover2016node2vec}, walk-based similarity scores that are the basis for link prediction \citep{liben2007link}, or heat kernels for graphs used for community detection \citep{kloster2014heat} and node ranking \citep{chung2007heat}.
To apply these methods, it is necessary to obtain a transition matrix that formalizes the underlying stochastic process in the observed graph.
This is trivial when we have full information on repeated interactions in the graph, which enables us to estimate transition probabilities between nodes based on relative frequencies of observed interactions.
However in real-world settings we are often confronted with incomplete data, which turns the construction of the transition matrix into an \emph{inference problem} that we need to address to obtain reliable results.

In this work, we consider situations where we have access to a possibly incomplete set of observed repeated interactions between a set of nodes.
Such data can be represented as weighted graph, where the weight of an edge corresponds to the number of observed interactions between a given node pair.
For such weighted graphs, it is common to define transition probabilities proportional to the edge weights.
From a statistical inference point of view, this method to infer the transition matrix based on observed interactions corresponds to a frequentist approach that uses a maximum likelihood estimation.
When few observations are available, this simple approach suffers from overfitting: 
On the one hand, unobserved interactions translate to zero transition probabilities even though transitions may actually be possible in the underlying graph. 
On the other hand, those interactions that were observed are likely to translate to overestimated transition probabilities.
This generates a large variance in inferred transition probabilities that can severely distort the results of downstream network analysis and graph learning tasks.

\begin{figure}[t]
    \centering
  \includegraphics[width = 0.75\textwidth]{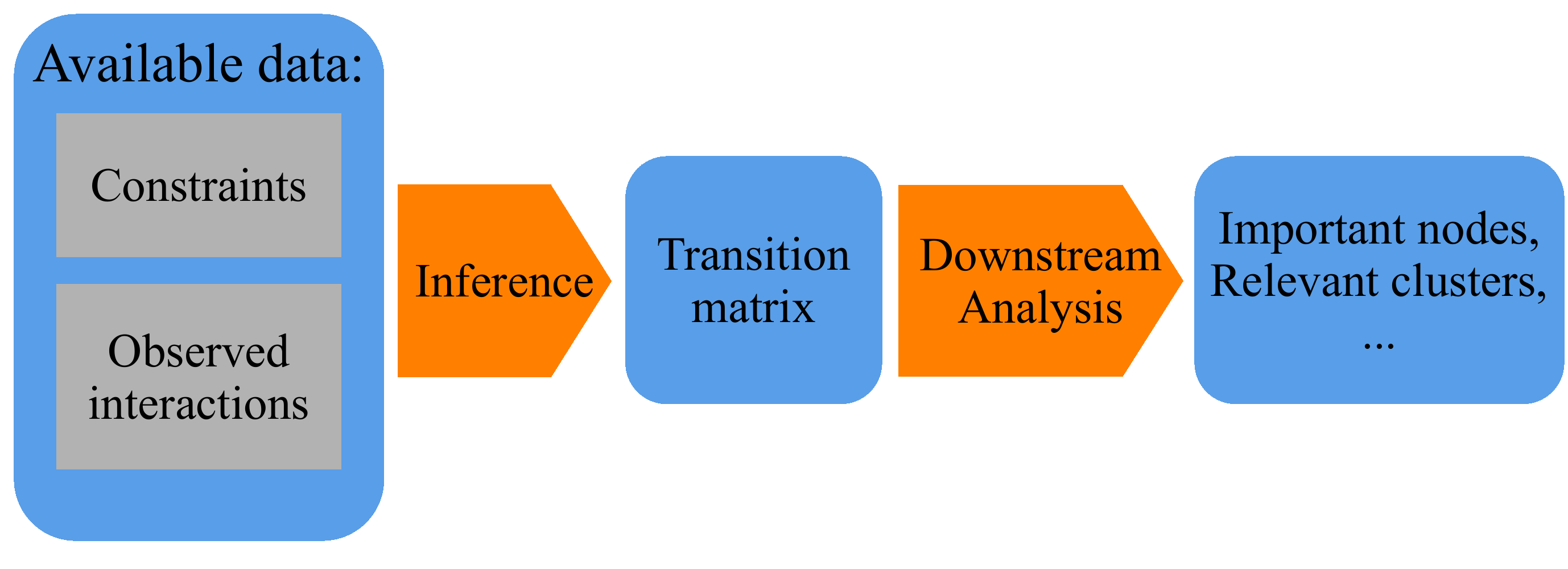}
  \caption{
    Steps in the analysis of network data.
    The first step is the inference of the transition matrix from the network data.
    In the second step, the transition matrix is used to detect the important nodes, relevant clusters and other interesting features of the system.
    The goals of our work are to extend the network inference to use the topological constraints, to investigate the robustness of the new method to partial knowledge of the constraints, and to investigate whether this improves downstream analysis tasks.
}
  \label{fig:infographics}
\end{figure}

However, repeated interactions are often not the only information that we have about the networked systems that we study.
We often have additional knowledge about topological constraints that determine which of the interactions are theoretically possible and which others are not.
For example, consider a transportation system, where observed interactions represent passengers travelling between connected stations, or a social network, where interactions represent messages transferred between users, or a Web graph, where interactions represent users clicking on a hyperlink between two Web pages.
In the first case, the movement of passengers is constrained by the available transportation infrastructure, in the second, the spreading of information is constrained by existing social connections, while in the third case, available hyperlinks constrain the possible clickstreams of users.
Such constraints can limit the number of parameters of the model that we seek to infer.
They can thus help us to address the reliable inference of transition matrices and improve the results of downstream analyses.
Therefore, in this paper, we address the following research questions:
\begin{enumerate}
    \item[\bf Q1] How can we use information on topological constraints to improve the inference of transition matrices from incomplete data on repeated interactions captured in weighted graphs?
    \item[\bf Q2] How is the inference of transition matrices influenced by a \emph{partial} knowledge of the underlying topological constraints, which is often the case in real-world settings?
    \item[\bf Q3] To what extent can our proposed approach to include topological constraints in the inference of transition matrices improve the performance of downstream network analysis tasks like node ranking and community detection?
\end{enumerate}
The remainder of this article is structured as follows:
In \cref{sec:background}, we formally define the inference problem that we address in our work and we introduce two methods that are commonly used to infer transition matrices without leveraging topological constraints, namely maximum likelihood estimation and a ``naive'' Bayesian approach.
Addressing Q1, in \cref{sec:BaCon} we introduce BaCon, a Bayesian approach that uses a topological prior to improve the inference of transition matrices in incomplete data on repeated interactions in a graph.
In \cref{sec:datasets}, we introduce the datasets and the experimental setup that we use to evaluate our method.
We next compare BaCon to the methods introduced in \cref{sec:background} (i.e. frequentist and naive Bayes approach) that do not use information on topological constraints.
In \cref{sec:results:inference}, we evaluate the extent to which the inclusion of the topological constraints improves the network inference, and address Q2, observing the effects of partial knowledge of the constraint.
In \cref{sec:results:analysis}, we address Q3 and explore whether the effects of the network inference carry over to the network analysis results.
In \cref{sec:results:skewed}, we show how inference of diverse examples of real-world networks can be further improved with an appropriate choice of prior parameters.
In \cref{sec:relatedWork} we discuss how our work complements existing network analytic methods in the field. 
We finally summarize our conclusions and outline future work in \cref{sec:conclusion}.

The results of our study show that (i) the inclusion of topological constraints considerably improves the inference of transition matrices in network data, and (ii) that this improved inference translates to an increased accuracy for downstream network analysis tasks.
Our work highlights the importance of treating the construction of network models from (partially observed) interactions as an inference problem that can be addressed using Bayesian statistics.
Our results further suggest that constraints for the topology of interactions in complex systems should be given more attention as their inclusion can significantly improve modelling accuracy, especially in the limit of small data.
Given the importance of network inference in partially observed or noisy data we expect our work to be of interest for a broad interdisciplinary community.
An implementation of our method is available as an Open Source project~\citep{zenodo_BaCon}.

\section{Background}

\newcommand{\sampledEdges}{\mathcal{E}}
\newcommand{\Tgt}{\mathbf{T}_\text{gt}}
\newcommand{\Tmat}{\mathbf{T}}

\label{sec:background}
\begin{figure}[!ht]
    \centering
  \includegraphics{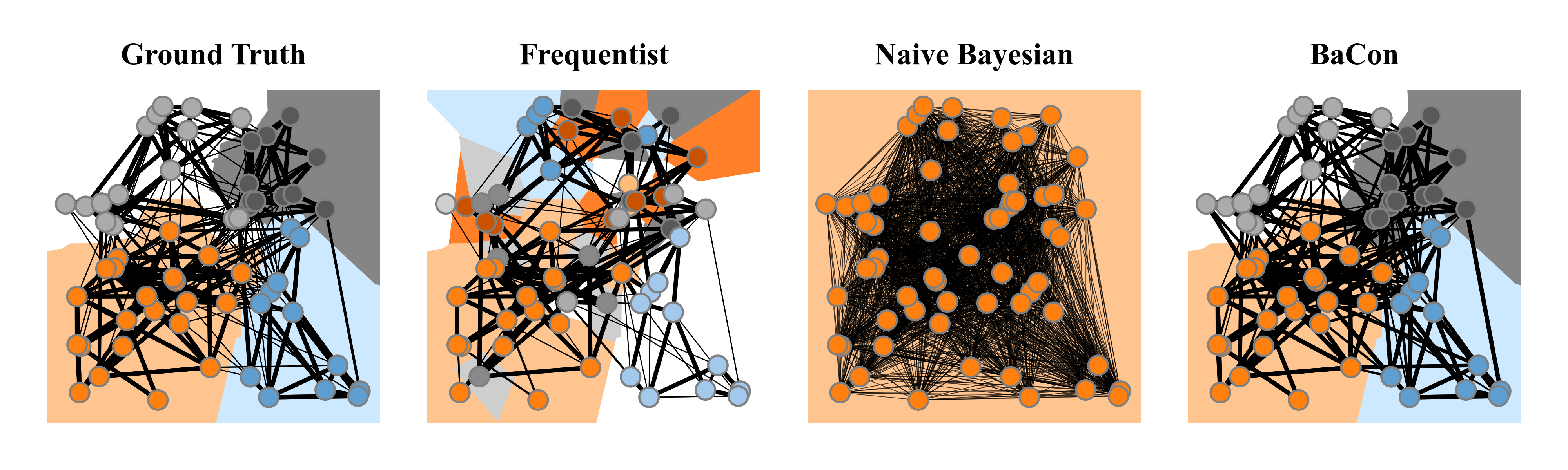}
  \caption{
  Illustration of methods to infer transition matrices from repeated interactions on a downstream task of clustering.
  The first panel shows the ground truth clusters encoded in the transition probabilities of a geometric random network with Euclidean metric in which we observe small set of interactions. 
  We use three different inference methods to construct a transition matrix and detect communities using InfoMap.
  The other panels show the clusters detected using the transition matrix inferred with a frequentist, naive Bayesian, and our approach - BaCon. 
  BaCon finds clusters that are closer to the ones detected from the ground truth matrix. 
}
  \label{fig:teaser}
\end{figure}
This section introduces the mathematical formalism for the inference of transition matrices from observations on repeated interactions and outlines two existing strategies for addressing it.
We assume that we observe a multiset of pairwise interactions $(i,j)$ between nodes $i,j \in V$, we denote it with $\sampledEdges$ and indicate the number of times the interaction $(i,j)$ occurred with $n_{ij}$.
In addition, we are given a directed graph $G=(V,E)$ that represents the topological constraints, i.e., which edges can be observed and which are impossibile. 
Given these sources of data, our goal is to infer the transition matrix $\Tmat$ that best reflects the actual transition probabilities in the system.

The simplest option is to assign the transition probabilities in a frequentist way, i.e. proportionally to the observation counts $n_{ij}$ and thus $\pi_{ij} = {n_{ij}}/{\sum_{k\in V} n_{ik}}$.
This corresponds to a maximum likelihood estimation (MLE) of a multinomial distribution $p(j|i)$.
The advantages of this approach are its simplicity and good performance when we have a data set that is sufficiently large considering the space of possible interactions.
In fact, this method is so common that it has become a standard ``preprocessing'' step, which is rarely even mentioned as a method to ``infer'' a graph model from observational data.
However, this simple method can require a large number of observations to assign non-zero probabilities to all edges that are possible based on the underlying topology of the system.
In a nutshell, when few observations are available, it is difficult to determine whether an edge with probability zero is not possible or whether it has not been observed yet.
From a machine learning point of view, the erroneous interpretation of edges with zero probability as evidence of the absence of the associated edge corresponds to an \emph{overfitting} of the weighted graph model.

An alternative to a frequentist approach is to use a Bayesian one. 
The naive Bayesian approach addresses the issue of overfitting by recording the distribution of parameters $\pi_{ij}$ for a given data set.
For every node $i$, we organize the parameters $\pi_{ij}$ in vectors $\vec{\pi}_i = (\pi_{ij})_{j \in V}$.
A priori, we would assign, e.g., a uniform prior over the space of transition probabilities: $p((\vec{\pi_{i}})_{i\in V}) = const. = \prod_{i\in V}\text{Dir}(\pi_{i} | \vec{\alpha}_i = \vec{1}_N),$ 
where $\text{Dir}$ denotes the Dirichlet distribution, $\vec{\alpha}_i$'s are its concentration parameters, and $\vec{1}_N$ is the $N$-dimensional vector with all components equal to $1$. 
This choice of parameters corresponds to the uniform distribution.
The naive Bayesian approach further uses Bayes' rule to update the prior distribution of transition probabilities:
$ p((\vec\pi_{i})_{i \in V} | \sampledEdges ) = {p( \sampledEdges | (\vec\pi_{i})_{i \in V}) p( (\vec\pi_{i})_{i \in V}) }/{ p(\sampledEdges) } $.
An advantage of this method is that unobserved edges are still modeled with non-zero probabilities.
However, this also introduces problems: 
First, since all transitions are modelled with non-zero probabilities, we cannot directly use sparse matrices, which complicates applications to large networks.
Second, and more importantly, in the typical case of networks with sparse topologies, a large amount of data is required to overcome the uniform prior on a fully connected graph.
From a machine learning point of view, this corresponds to an \emph{underfitting} of the weighted graph model.

In the first three panels of \cref{fig:teaser}, we use a toy example to illustrate the issues of frequentist and naive Bayesian network inferences. 
In the toy example, we consider a ground-truth transition matrix constrained to a random geometric graph topology (first panel). 
The network consists of four clusters expressed in higher transition probabilities between nodes within the same cluster.
We draw a small random sample of repeated interactions from the distribution of edge probabilities in the ground truth network.
We apply the frequentist (second panel) and the naive Bayesian method (third panel) to infer a transition matrix.
We further visualize the result of the popular community detection technique InfoMap on the resulting transition matrix. 
The frequentist approach detects many spurious communities because of its overfitting issue. 
The Bayesian approach detects a single community because the observed data is insufficient for overcoming the prior.

Neither the frequentist approach, nor the naive Bayesian, use the information on the constraints $G=(V,E)$. 
Addressing this issue, in the next section, we formally introduce a Bayesian method that leverages topological constraints (BaCon) that are often known in real-world networked systems.
In the fourth panel of \cref{fig:teaser}, we show a representative example that illustrates how inclusion of such constraints in the network inference improves the detection of clusters.

\section{BaCon: Bayesian Constrained Network Inference} 
\label{sec:BaCon}

In this section, we show how the knowledge of topological constraints can be used as a prior to infer transition probabilities from observations of interactions (Q1).
We formally introduce a Bayesian method (BaCon) that leverages topological constraints that are often available for real-world networked systems.

We assume that we are given a set of nodes $V$ of size $N$, and a set of possible edges $E \subset V \times V$ that represent a topological constraint for our inference task.
I.e. we assume that an interaction between node $i$ and node $j$ cannot occur if $(i,j) \notin E$.
For each node $i$, we denote its possible successors as $S(i)$.
We also assume that we are given a data set $\sampledEdges$ of interactions that were observed in the system.
Our goal is to infer the best directed weighted graph model from this data set.
The transition probabilities $p(j|i) = \pi_{ij}$ from $i$ to its successors $j \in S(v)$ satisfy:
\begin{equation}
    \sum_{j\in S(i)} \pi_{ij} = 1, \forall i,j: 0 < \pi_{ij} < 1 
    \label{eq:probabilities}
\end{equation}
We are interested in the probability density of parameters $\pi_{ij}$.
We again organize them in vectors $\vec{\pi}_{i}$, similarly to the naive Bayesian approach introduced in the previous section, but this time only over the possible successors: $\vec{\pi}_{i} = ( \pi_{ij})_{j \in S(i)}$.
We capture the information about impossible transitions using a Dirac delta function as prior for the $\pi_{ij}$.
Since those transitions can not occur in the data, their probability distribution will not change.
For the possible transitions from a node $i$, we assume a uniform prior over the parameter space:
\begin{equation}
\label{eq:prior}
p(\vec\pi_i|G) = \text{Dir}\left(\vec{\pi}_{i} \Big| \vec{\alpha}_{i} = \alpha \vec{1}_{|S(i)|}\right)
\end{equation}
where, for the uniform distribution,  we make a choice of $\alpha = 1$.
The choice of $\alpha<1$ means that the prior distribution of the probabilities is more skewed, and the choice of $\alpha>1$ means that the prior distribution of the probabilities is more peaked around the value $1/|S(i)|$.
In the following we denote all transition probabilities as $\pi := (\vec\pi_i)_{i\in V}$.
The likelihood of observing data $\sampledEdges$, which contains $n_{ij}$ observations of an edge $(i,j)$, is given by the multinomial distribution:
\begin{equation}
    p(\sampledEdges|\pi,\alpha,G) = Z \prod_{i} \prod_{j \in S(i)} \pi_{ij}^{n_{ij}}
\end{equation}
where $Z$ denotes the number of permutations of observations.
We use Bayes' rule to update our a priori distribution after observing data $\sampledEdges$:
\begin{equation}
    p(\pi | \sampledEdges, \alpha, G ) =
    \frac{
    p( \sampledEdges | \pi,\alpha, G )
    \times
    p(\pi | \alpha, G )
    }{
    p( \sampledEdges  | \alpha, G )
    }
\end{equation}
As the multinomial and the Dirichlet distributions are conjugate distributions, the posterior is also a Dirichlet distribution with parameters 
$ \vec\alpha_{i} = \vec\alpha_{i}^0 + \vec n_{i},$
where $\vec\alpha_{i}^0$ denotes the a priori concentration parameters, and $\vec n_{i} := (n_{ij})_{j\in S(i)}$.
The posterior
$
p(\vec\pi_{i} | \sampledEdges) = \text{Dir}\left(\vec\pi_i \Big| \vec\alpha_i =\alpha \times \vec{1}_{|S(i)|} + \vec n_{i}\right)
$
defines an ensemble of parameters consistent with the observations.
From this ensemble we can compute the expected value of the transition matrix as:
$$ \mathbf{E} \left[ T_{ij} \right] = \frac{\alpha_{ij}}{\sum_{k} \alpha_{ik}} $$
We will use the expected value of the transition matrix in our experiments where we compare the outlined inference method versus the alternatives.

We showed a method to infer the transition matrix $\Tmat$ from observed interactions between system elements, using the topological information as a key ingredient of the prior distribution, which provides an answer to the first research question. 
We highlight that the method has one free parameter $\alpha$, which governs how peaked or skewed is the prior distribution of the transition probabilities.
In our experiments, we chose $\alpha = 1$ which corresponds to the uniform prior.
However, we experimentally show the effect of different choices of $\alpha$, and how we can use it to leverage prior knowledge on the distribution in \cref{sec:results:skewed}.

\section{Datasets}
\label{sec:datasets}
In this section, we describe the synthetic and empirical datasets that we use to test BaCon and investigate the research questions Q2 and Q3.
For the synthetic data-sets, we generate the ground truth transition matrix using three different random graph models and two task-specific probability generation processes.
We further use five real world datasets from ecology, neurology, information systems, transportation systems and technical systems. 

\subsection{Synthetic Datasets}
\label{sec:datasets:synthetic}

In this section we describe the procedure that we use to generate the synthetic data.
First, we generate the underlying topology $G=(V, E)$, then, based on this topology, we randomly generate transition probabilities of a ground truth transition matrix $\Tgt$, and finally, we use the transition matrix to generate the synthetic interactions $\sampledEdges$.
This synthetic generation process is necessary to ensure that the target pattern (e.g. the ground truth clustering or the ground truth ranking) cannot be fully recovered only from the graph topology $G=(V,E)$, and thus that the ground truth probabilities are necessary for the downstream task.

\newcommand{\sigmaRGG}{\sigma}

\paragraph{Underlying topology}
To generate the underlying topology for our synthetic experiments, we first generate an undirected network using one of three different generative models: (1) an Erd\H{o}s-R\'enyi $G(N,M)$ random graph model, (2) soft random geometric graphs with Euclidean metrics of latent spaces, and (3) soft random geometric graphs with hyperbolic metrics of latent spaces. 
The Erd\H{o}s-R\'enyi $G(N,M)$ model (where $N = 500$ is the number of nodes and $M = 5000$ the number of edges) is one of the simplest random graph models. 
It generates networks with the small distance between pairs of nodes that characterizes real networks, but without their degree heterogeneity, high clustering values, and modular structures.
In contrast, soft random geometric graphs have high clustering and display the emergence of modular structures. 
They generate networks with different degree distributions depending on the metric of the latent space. 
In our experiments, we use both a Euclidean metric, which generates networks with high clustering but uniform degree distributions, and a hyperbolic metric, which generates networks with high clustering and power-law degree distributions \citep{krioukov2010hyperbolic}.
To produce a geometric graph, we scatter $N = 500$ nodes in a two dimensional latent space. 
For the Euclidean metric, we uniformly scatter nodes in a $(0,1)\times(0,1)$ square.
For the hyperbolic metric, we uniformly scatter nodes in a hyperbolic disk of radius $R=1$ and constant Gaussian curvature $-\zeta^2$ with $\zeta = 1$.
We connect nodes $i$ and $j$ with probability $p((i,j)) =  \text{exp}(-{d_{ij}}/{\sigmaRGG}) $, where $d_{ij}$ is the pairwise distance of nodes in the latent space and $\sigmaRGG = 0.1$ is the scale parameter.
Large values of $\sigmaRGG$ indicate a higher probability of connecting to distant nodes, leading to denser networks.
Low values of $\sigmaRGG$ favor connections with closer nodes and lead to sparser topologies.
Having generated the network, we add an edge from a random node of every connected component to a random node in the largest connected component, thus enforcing the connectedness of the network.
Finally, we make the network directed by converting each undirected edge $(i,j)$ to two directed edges $(i,j)$ and $(j,i)$, thus obtaining $G = (V,E)$ that is the underlying topology.
For this choices of parameters, Erd\H{o}s-R\'enyi networks contain around $10000$ directed edges, Euclidean networks produce around $12000$ directed edges and hyperbolic networks produce around $4000$ directed edges.
Obtaining the network constraint from a connected undirected network ensures that random realizations of the synthetic transition matrix has no sinks. 

\paragraph{Ground truth transition matrix}
Next, we set the probabilities on the underlying topology, thus defining the ground truth model. 
To generate the ground truth transition matrix $\Tgt$, we generate the transition probabilities $\pi_{ij}$ of outgoing edges for each node $i$ from $(\pi_{ij})_{j \in S(i)} \sim \text{Dir}\left(  \vec\alpha_i \right)$, with two different choices of $\alpha_i$.

For the experiments on transition matrix inference and on node ranking, we choose a uniform distribution: $\vec\alpha_i = \vec1_{S(i)}$.
However, to ensure that we do not produce networks with a single community, we employed the following procedure.
We first artificially group the nodes, and choose a larger value for the component $\alpha_{ij}$ of $\vec\alpha_i$ when $j$ is in the same community as $i$ ($\alpha_{ij} =10$) then when it is not ($\alpha_{ij} =1$).
We group the nodes of Erd\H{o}s-R\'enyi graph by uniformly placing nodes in three groups; in the case of random Euclidean graph, we cut the latent space with a random horizontal and a random vertical line, and assign nodes to the same group if they belong to the same partition of the latent space; in the case of random hyperbolic graph, the procedure is the same, except that we cut the space with two geodesics.
By grouping nodes in this way, we ensure that the topology of the geometric graphs holds some information about the cluster structures, but that the probabilities are not irrelevant for detecting the cluster structure.

\paragraph{Edge sampling}
In the third step, we generate a sample of interactions $\sampledEdges$.
Since we do not expect the starting nodes to be observed with uniform frequency, we assign random ``starting probabilities'' to different nodes.
We draw the starting probabilities from $\text{Dir}\left( \vec\alpha = \vec{1}_{N} \right)$.
We choose the starting node $i$ as a multinomial draw of the starting node probability distribution.
We choose the successor $j$ as a multinomial draw of the probability distribution of successors.
The sampling simulates the observation of interactions in a networked system.

\subsection{Empirical Datasets}
\label{sec:datasets:empirical}
In this section we describe the empirical weighted networks that we used to evaluate the methods.
The five real-world empirical networks come from different domains.
 
The first network captures an \emph{ecosystem}, which is the food chain of Florida bay \citep{veraszto2020whole}.
It contains $2106$ edges capturing carbon exchange between $128$ species.
An entry $A_{ij}$ of the adjacency matrix represent the biomass exchanged from species $i$ to species $j$ (i.e. $j$ ate $A_{ij}$ quantity of $i$).
We assume that the non-zero values of the adjacency matrix define the possible edges.

The second network, denoted as \emph{neural}, in the domain of neuroscience, captures the connectome (i.e. the map of neural connections) of a larva of a simple marine worm \citep{ulanowicz2005network}.
The topology is defined with $11437$ axons connecting $2728$ neurons.
The entry $A_{ij}$ of the adjacency matrix is the number of synapsis connecting neuron $i$ to neuron $j$.

The third network is the Wikipedia web graph used to play the \emph{wikispeedia} game in \citep{west2012human}. 
In the game, players have to find a short path between two pages in the graph.
The topological constraint is defined by the $239 764$ hyperlinks that the players could use to navigate through $4592$ Wikipedia pages.
An entry of the adjacency matrix $A_{ij}$ is the number of times users clicked on the hyperlink from $i$ to $j$ while playing the game.

The fourth network records \emph{flights} between US airports. 
The data~\citep{FLdata} captures $286 810$ passenger itineraries between $175$ US airports as recorded in 2014. 
An entry of the adjacency matrix $A_{ij}$ is the number of passengers that flew from $i$ to $j$, and we assume that the connection does not exist if none of the observed passengers used it.

The fifth network is the central Chilean \emph{power-grid} \citep{kim2018depth}.
The topological constraint is defined from $444$ connections between $347$ stations.
The entry of the adjacency matrix $A_{ij}$ is the capacity of the connection measured in Kilovolts.
Although the \emph{powergrid}, \emph{neural}, and \emph{ecosystem} data sets do not quite fit in our setting (e.g. connection voltage does not determine how often we observe the connection), we can still use them to test the inference methods on a diverse set of real-world networks.

Each of the empirical networks defines the adjacency matrix $\mathbf{A}$. 
We define the underlying constraint $G=(V,E)$ as a graph of non-zero weighted edges $(i,j)\in E \Leftrightarrow A_{ij} > 0$.
The ground truth transition matrix $\Tgt$ is obtained by normalizing rows of the adjacency matrix $\mathbf{A}$.
We simulate observations by sampling edges from the adjacency matrix $\mathbf{A}$ of the corresponding weighted network. 
We put all edges $(i,j)$ in a single bin and draw them with probability proportional to $A_{ij}$.
The sampling simulates an input of interactions observed on a networked system.
Since the sampling introduces variability on the observed edges, we run the experiments multiple times.

\section{Effects on Inference of Weighted Graphs}
\label{sec:results:inference}

We now explore the effect of using topological constraints in the inference of transition matrices.
We compare the proposed method BaCon, which uses information on the topological constraint, to the frequentist and the naive Bayesian approaches introduced in \cref{sec:background}, which do not incorporate topological information.
At the end of the section, we investigate the research question Q2 and explore how partial knowledge of topological constraints influences the inference.

\paragraph{Experimental setup}
The input data consists of a sample of observed edges $\sampledEdges$ and a given graph constraint $G$ as described in \cref{sec:datasets}.
The target variable is the ground truth transition matrix $\Tgt$, which we construct as explained in \cref{sec:datasets}.
We use the data to infer a transition matrix $\Tmat$ using each method and compare those transition matrices to the ground truth $\Tgt$.
We quantify the errors in estimating the matrix entries by computing the Frobenius norm of the difference between the inferred transition matrix and the ground truth transition matrix $\|\Tgt - \Tmat\|$.
Since the Frobenius norm of a matrix $\mathbf{X}$ is defined as $ \|\mathbf{X}\| = (\sum_{ij} X_{ij}^2)^{1/2}$, computing the Frobenius norm of the difference between transition matrices is equivalent to computing the mean square error of the transition probabilities inferred by the models. 
The worst value for $\|\Tgt - \Tmat\|$ is $2N$ where $N$ is the number of nodes in $G$; the best is $0$, which indicates a perfect match between the inferred matrix and the ground truth transition matrix.
We evaluate the performance of methods for different sizes of the sample $\sampledEdges$ and we plot how the error of the inferred transition matrix depends on the number of sampled edges $\sampledEdges$.
We run the experiment $100$ times for each sample size.
In all figures, we represent the average error of the frequentist method with the dotted black line, the dashed dark-blue curve shows the average error of the naive Bayesian method.
The average error of BaCon is represented with the solid orange curve.
As a reference point, we drew the light-gray dashed horizontal line, which represents the average difference between the ground truth transition matrix and the transition matrix constructed based on the unweighted topology, which does not depend on the sampled edges.
We represent the variability in the results' distribution using vertical error bars that indicate the $95\%$ confidence intervals (specifically, the intervals between the $2.5$-th and the $97.5$-th percentiles).

\begin{figure*}[!ht]
     \centering
        \includegraphics{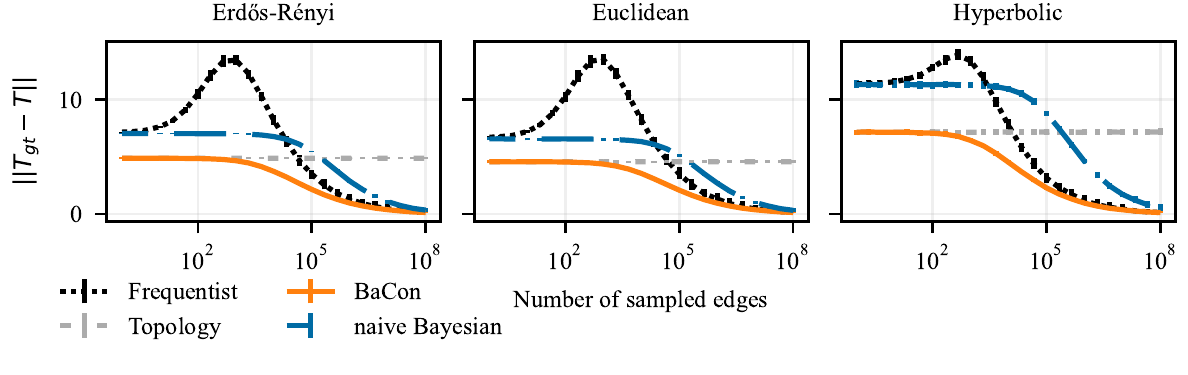}
        \caption{
            Effects of the use of constraints on the inference of a transition matrix in synthetic data.
            The left panel shows results on Erd\H{o}s R\'enyi graphs, the middle panel shows results on geometric Euclidean random graphs, and the right panel shows results on geometric hyperbolic random graphs. 
            We measure performance with the Frobenius distance between the transition matrix of the ground truth and that of the inferred model.
            Error bars represent the $95\%$ confidence interval..
            The inclusion of the constraints allows BaCon to recover the transition matrix more precisely than both the frequentist and the naive Bayesian approaches. 
        }
        \label{fig:inference}
\end{figure*}

\begin{figure}[!ht]
 \centering
     \includegraphics{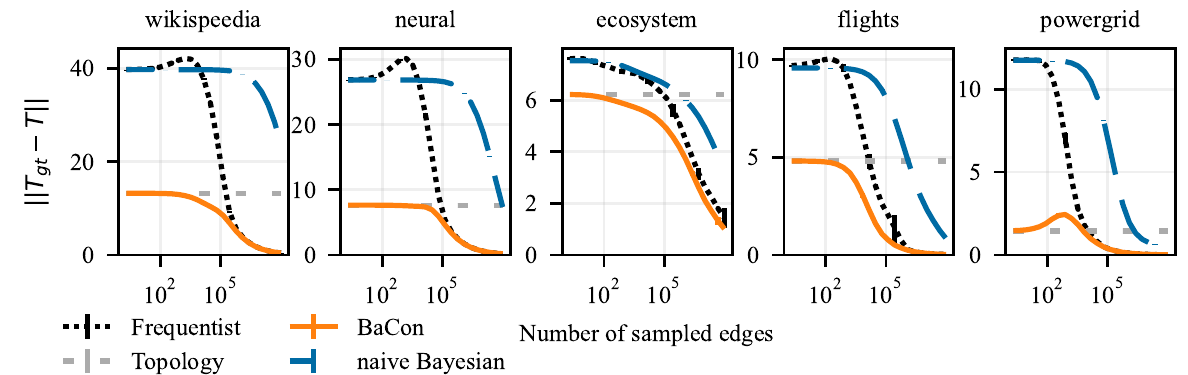}
    \caption{
        Effects of the use of constraints on the inference of a transition matrix in empirical data. 
        From left to right, the panels show the results for \emph{wikispeedia}, \emph{neural}, \emph{ecosystem}, \emph{flights}, and \emph{powergrid} networks.
        We measure performance with the Frobenius distance between the transition matrix of the ground truth and that of the inferred model.
        Error bars represent the $95\%$ confidence interval.
        In accordance with the synthetic results, the inclusion of the constraints allows BaCon to recover the transition matrix more precisely than both the frequentist and the naive Bayesian approaches. 
     }
    \label{fig:inference_empirical}
\end{figure}

\paragraph{Results}
We present our results on the dependency between the inference error ($y$-axis) and the sample size of interactions ($x$-axis) in \cref{fig:inference} for synthetic networks and in \cref{fig:inference_empirical} for empirical networks.
The results show that the inclusion of the topological constraint positively affects the inference of the transition matrix.
All three methods show an improvement in performance with the number of sampled edges and converge to the ground truth transition matrix for a large number of observations.
However, their behavior differs greatly in the intermediate and small data range. 
When few edges are available, the frequentist and naive Bayes perform similarly, and both do worse than the unweighted topology.
In the intermediate range, the two methods display very different behaviors.
The frequentist approach exhibits a peak in its error (particularly marked for the synthetic experiments), indicating that the method over-adjusts to the observations.
On the contrary, the naive Bayesian method steadily improves performance but at a very slow pace: due to the use of a fully connected prior (i.e. the lack of topological information) each observation only provides very little information for the inference.
As a result, the frequentist method surpasses the naive Bayesian when the number of observed edges increases despite its peak in error in the intermediate data range.
BaCon mediates between these two behaviors. 
Its inference begins from a better initial performance by relying on the constraint provided by the unweighted topology.
In the intermediate data range, BaCon's performance improves considerably faster than the naive Bayesian approach because, due to the topological constraints, it has less degrees of freedom.
The constraint also gives BaCon the necessary information to avoid overreliance on the observations, thus preventing the peak in error that characterizes the frequentist method.
BaCon steadily improves its performance when more data becomes available, converging to the ground truth transition matrix faster than the other approaches.

To give an idea of how large these improvements can be, an aspect which might be hidden by the logarithmic axes, we highlight the difference in the amount of data required by the different methods to reach a fixed mean square error of $2.5$ in synthetic data.
The frequentist method needed over three times more data than BaCon in the case of Erd\H{o}s-R\'enyi random graphs, over four times more data in Euclidean random graphs and double the amount of data in the case of hyperbolic random graphs.
For the naive Bayesian method the ratios are even more pronounced: it needed over twenty times more data than BaCon in the case of Erd\H{o}s-R\'enyi random graphs and Euclidean random graphs and over fifty times more data in the case of hyperbolic random graphs.

\paragraph{Discussion}
With these experiments, we have investigated the impact of including topological constraints on the inference of a transition matrix.
The results highlight that the inclusion of constraints provides a major improvement in data-efficiency.
We noticed both the overfitting of the frequentist and the underfitting of the naive Bayesian approaches introduced in \cref{sec:background}.
The inclusion of constraints provides an alternative source of information in the limit of small data sizes which prevents overfitting. 
It also reduces the degrees of freedom, which prevents underfitting.

It is worth mentioning that the Frobenius norm, which we used to evaluate the inferred matrix, is not the only matrix norm we tried. 
However, the other matrix norms showed the same trends, and we selected the Frobenius norm because of the intuitiveness of its connection to the mean square error. 
The results for the other matrix norms are available in the complementary Zenodo package \citep{zenodo_BaCon}
 
One potential threat to the validity of these results is that the constraint we used in the inference is identical to the constraint we used to generate the data.
However, assuming full knowledge of the constraints is unrealistic for real-world systems.
In the next section, we tackle this case to check whether missing information about constraints invalidates the BaCon approach.

\subsubsection*{Sensitivity to partial constraints}

We now investigate research question Q2 and test the robustness of our inference method against incomplete knowledge about the underlying topology.
We introduce noise to the constraints by adding spurious edges alongside those that capture the underlying topology.
We note that the addition of such spurious edges effectively corresponds to a loss of information on the constraint.
An addition of all possible edges corresponds to the naive Bayesian method with a fully connected prior introduced in \cref{sec:background}.
Aiming for an equal number of original and spurious edges, we add $10000$ uniformly random spurious directed edges to the Erd\H{o}s-R\'enyi random graphs, $12000$ to geometric Euclidean random graphs, and $4000$ to the geometric hyperbolic random graphs.

\begin{figure*}[!ht]
     \centering
        \includegraphics{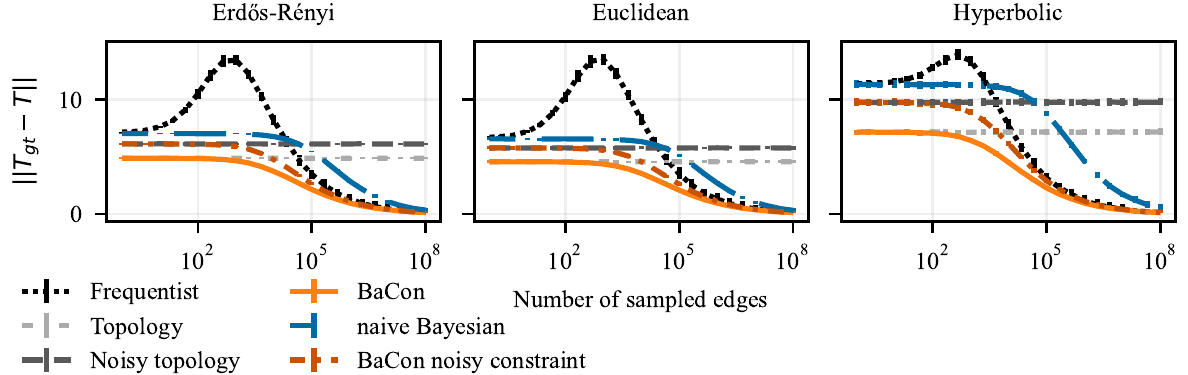}
        \caption{
        Effects of partial constraints on the inference of a transition matrix. 
        The left panel shows results on Erd\H{o}s R\'enyi graphs, the middle panel shows results on geometric Euclidean random graphs, and the right panel shows results on geometric hyperbolic random graphs. 
        We measure performance with the Frobenius distance between the transition matrix of the ground truth and that of the inferred model.
        Error bars represent the $95\%$ confidence interval.
        Despite the partial information on constraints, its usage benefits the network inference. 
        }
        \label{fig:inference_noisy}
\end{figure*}

The results in \cref{fig:inference_noisy} display that missing information on the constraints decreases the performance of BaCon in the small data regime, and that it needs more data to make up for that deficiency.
However, the method still outperforms the other two approaches, which do not use information about constraints at all. 
Answering Q2, this demonstrates that the proposed method BaCon is robust to missing information about topological constraints in complex systems.

\section{Effects on Downstream Analyses of Weighted Graphs}
\label{sec:results:analysis}

In this section, we investigate research question Q3 and measure the effects of the inference on the downstream tasks of node ranking and clustering.

\paragraph{Experimental setup} To evaluate the impact of the inference methods on downstream network analysis tasks, we perform experiments that build on the experimental setup from \cref{sec:results:inference}.
The input data again consists of the sample of edges $\sampledEdges$ and a graph constraint $G$, as described in section \cref{sec:datasets}.
From the input data, we infer the transition matrix using all three inference methods.
However, instead of evaluating the error of the inferred matrix $\Tmat$, in these experiments we perform downstream tasks of node ranking and cluster detection using the inferred transition matrix.
We use PageRank to compute a node ranking on $\Tmat$, and compare it to the corresponding PageRank node ranking obtained based on the ground truth transition matrix $\Tgt$.
We measure the performance using the Kendall-$\tau$ correlation between the two rankings; a perfect match of the rankings leads to $\tau = 1$, uncorrelated rankings lead to $\tau \approx 0$, whereas the reverse ranking is captured by $\tau = -1$.
To detect clusters, we apply the random-walk based community detection algorithm Infomap \citep{rosvall2008maps}.
This method is a convenient choice for our analysis as it naturally detects clusters in directed networks along with the optimal number of clusters.
We measure the performance by computing the adjusted mutual information (AMI) between the clusters detected in the inferred transition matrix $\Tmat$ and the clusters detected using the ground truth transition matrix $\Tgt$.
A perfect agreement has an AMI score equal to $1$, completely unrelated assignments have AMI of $0$.
We run $100$ independent experiments for each sample size and each task.
We plot the average performance of downstream tasks as a function of the size of the edge sample $\sampledEdges$. 
To represent the variability of the performance, we show the $95\%$ confidence intervals as the error bars.
The average performance using the frequentist inference is presented with the dotted black curve, the dashed blue is for the naive Bayesian method, and the solid orange is showing the performance of BaCon.
The dashed light-gray horizontal line, which we show as a reference point, again represents the average performance obtained using only the unweighted topology.

\begin{figure}
 \centering
    \includegraphics{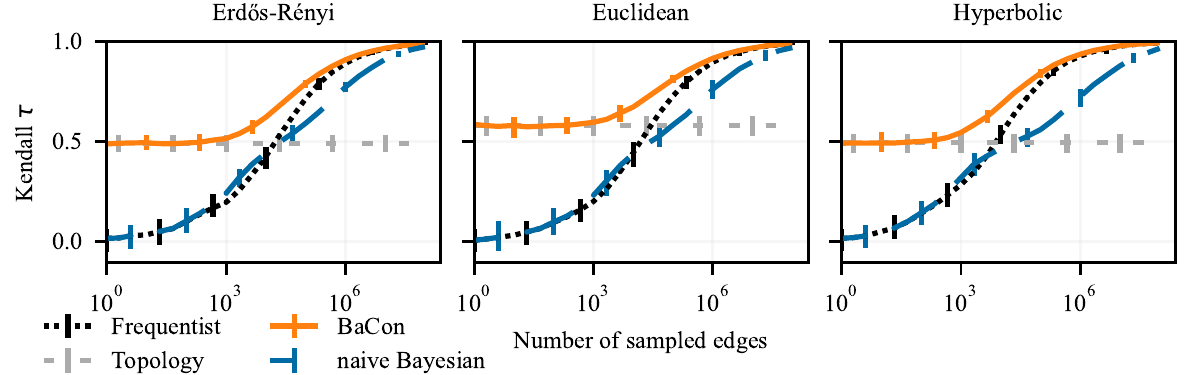}
    \caption{
        Effects of constraints on the downstream task of node ranking. 
        The left panel shows results on Erd\H{o}s R\'enyi graphs, the middle panel shows results on geometric Euclidean random graphs, and the right panel shows results on geometric hyperbolic random graphs. 
        We measure performance with the Kendall $\tau$ correlation between the ranking obtained from the ground truth and the inferred transition matrices.
        Error bars represent the $95\%$ confidence interval.
        The positive impact of constraints on network inference translates to the node ranking task. 
    }
    \label{fig:fig_PR_synth}
\end{figure}

\begin{figure}[!ht]
 \centering
     \includegraphics{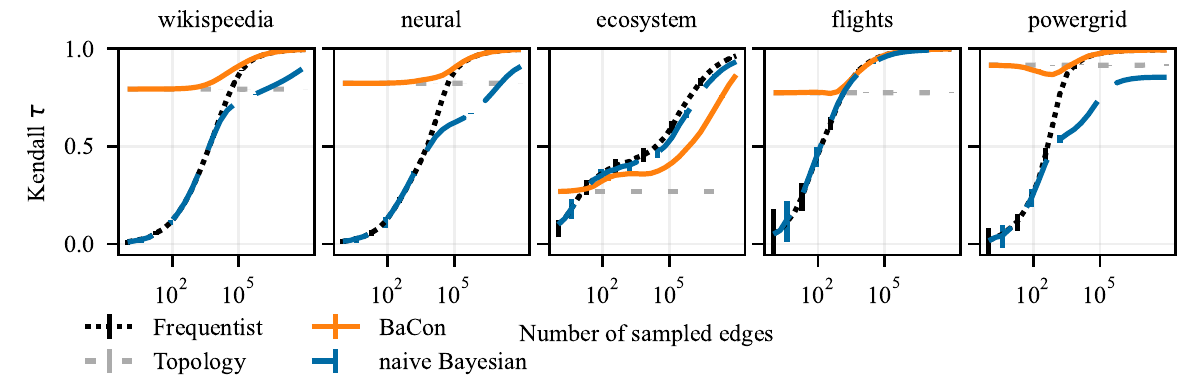}
    \caption{
            Effects of constraints on the downstream task of node ranking. 
             From left to right, the panels show the results for \emph{wikispeedia}, \emph{neural}, \emph{ecosystem}, \emph{flights}, and \emph{powergrid} networks.
            We measure performance with the Kendall $\tau$ correlation between the ranking obtained from the ground truth and the inferred transition matrices.
            Error bars represent the $95\%$ confidence interval.
            The positive impact of constraints on network inference translates to the node ranking task. 
    }
    \label{fig:PageRank_empirical}
\end{figure}

\paragraph{Node ranking} We present the results of the node ranking experiments on synthetic networks in \cref{fig:fig_PR_synth}, and on the empirical networks in \cref{fig:PageRank_empirical}. 
Except for the \emph{ecosystem} dataset, the inclusion of constraints in network inference improves the ranking of nodes. 
Both the frequentist and the naive Bayesian methods start from a performance of $\tau = 0$ in the limit of very small data sizes, and need considerable data to reach the performance of the unweighted topology.
On the other hand, BaCon defaults to the performance of the topology in the limit of small data. 
BaCon performs at least as good as the unweighted topology, with the only exception of the \emph{powergrid} dataset.
However, also in the \emph{powergrid} dataset, BaCon performs better than the frequentist and the naive Bayesian approaches.

\begin{figure}
     \centering
        \includegraphics{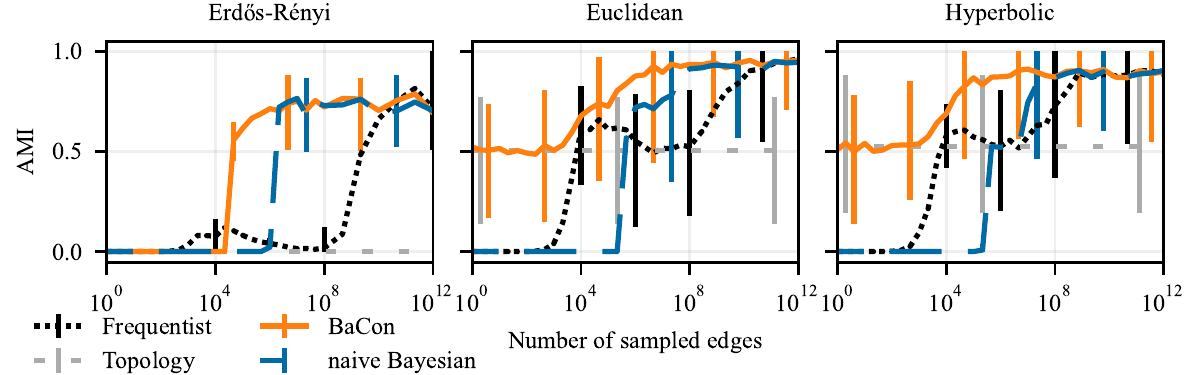}
        \caption{
            Effects of partial constraints on the downstream task of clustering. 
            The left panel shows results on Erd\H{o}s R\'enyi graphs, the middle panel shows results on geometric Euclidean random graphs, and the right panel shows results on geometric hyperbolic random graphs.
            We measure performance with the Adjusted Mutual Information (AMI) between the clusters obtained from the ground truth and the inferred transition matrices.
            Error bars represent the $95\%$ confidence interval.
            The positive impact of constraints on network inference translates to the clustering task. 
        }
        \label{fig:clustering}
\end{figure}

\begin{figure}
     \centering
        \includegraphics{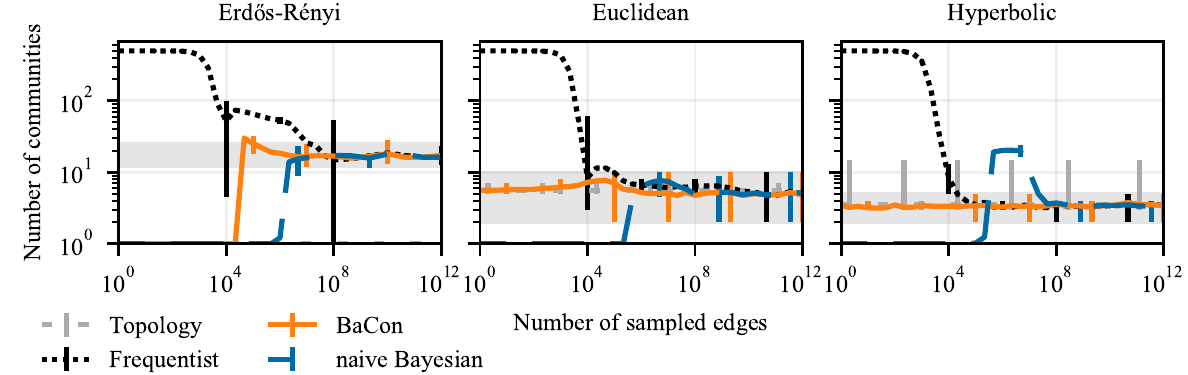}
        \caption{ 
            Number of detected clusters by each method for a given number of sampled edges. 
            The left panel shows results on Erd\H{o}s R\'enyi graphs, the middle panel shows results on geometric Euclidean random graphs, and the right panel shows results on geometric hyperbolic random graphs.
            We show the number of communities detected by each method against the horizontal band indicating the interval between $2.5$-th and the $97.5$-th percentiles of the ground truth number of communities (shaded gray area).
            Error bars represent the $95\%$ confidence interval.
            The positive impact of constraints on network inference translates to the clustering task. 
        }
        \label{fig:number_of_clusters_syn}
\end{figure}

\begin{figure}[!ht]
 \centering
     \includegraphics{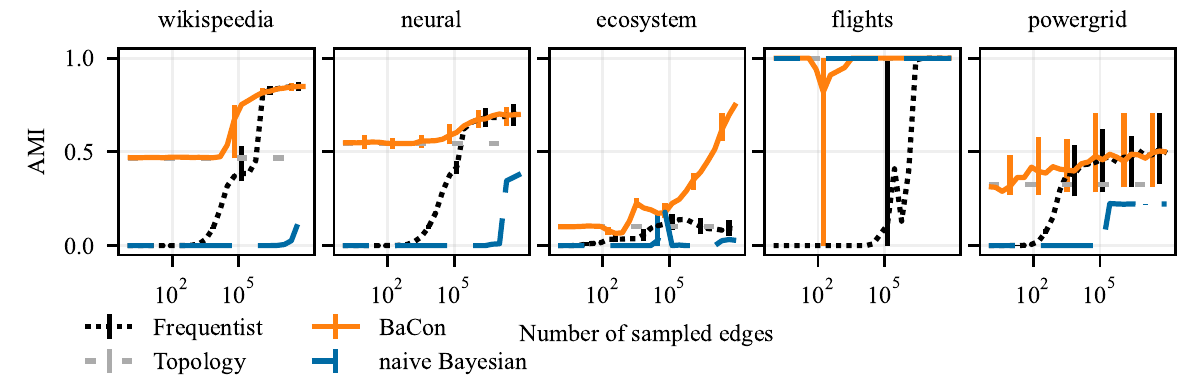}
    \caption{
        Effects of constraints on the downstream task of clustering. 
         From left to right, the panels show the results for \emph{wikispeedia}, \emph{neural}, \emph{ecosystem}, \emph{flights}, and \emph{powergrid} networks.
        We measure performance with the Adjusted Mutual Information (AMI) between the clusters obtained from the ground truth and the inferred transition matrices.
        Error bars represent the $95\%$ confidence interval.
        The positive impact of constraints on network inference translates to the clustering task. 
    }
    \label{fig:clust_empirical}
\end{figure}

\paragraph{Cluster detection} We present results of experiments on cluster detection in synthetic data in \cref{fig:clustering} and in empirical data in \cref{fig:clust_empirical}.
These results show that the constraints also improve the detection of clusters.
The \emph{flights} dataset might appear as an outlier, where the naive Bayesian inference outperforms the other methods. 
However, this is merely due to the fact that the \emph{flights} dataset exhibits only a single community, which is what the naive Bayesian method always predicts a priori.
Although the variance of the performances is larger compared to the previous task, we see similar behaviour as in the node ranking experiments.
While the frequentist and the naive Bayesian approaches have AMI of zero in the limit of small data sizes (indicating no correlation between the identified and ground truth clusters), BaCon defaults to the performance of the unweighted topology.
We further inspect this behaviour in \cref{fig:number_of_clusters_syn}, where we show the number of detected clusters in the synthetic graphs introduced in \cref{sec:datasets:synthetic}.
We compare them against the number of clusters detected in the ground truth transition matrix, indicated with the gray horizontal band representing the $95\%$ confidence interval.
The naive Bayesian method needs considerably more data to detect more than one cluster, while the frequentist method generally overestimated the number of clusters.
This means that, although their performance in terms of AMI is similar, they err differently.



\paragraph{Discussion}
With the experiments presented above, we investigated the consequences of using topological constraints on downstream network analysis tasks, i.e. research question Q3 from \cref{sec:introduction}.
The experiments show that the inclusion of topological constraints in the inference generally improves the performance of downstream network analysis tasks.
The results also show that the underfitting of the naive Bayesian method and the overfitting of the frequentist method carry over to downstream tasks.
Not surprisingly, both methods initially require a sufficient number of observed edges to reach the performance obtained by the unweighted topology alone.
This underlines the dilemma that practitioners face when they are confronted with a dataset of edge observations and information of the topology: Which of the two sources will lead to a more reliable analysis of the underlying network?
Instead of choosing between the two sources of information, BaCon allows to include both. 
As a result, the downstream analysis is as good as for the unweighted topology in the limit of small number of observation, and generally at least as good as the other two methods.

The two outliers, the \emph{ecosystem} and the \emph{powergrid} datasets, require special consideration.
Although the inclusion of constraints improved the inference and the performance of the downstream task of cluster detection, they showed different behaviour in the node ranking task.
In the \emph{powergrid} dataset, although we see that BaCon outperforms the frequentist and naive Bayesian methods, which means that the inclusion of topological constraints improves the inference, we observe that in some cases the unweighted topology leads to better node ranking than the transition matrix inferred with BaCon.
In other words, for some data sizes, including observations deteriorates the ranking.
The \emph{powergrid} dataset represents an outlier because most of the weights have the same value since they represent voltages.
For such a peaked distribution of the weights, the uniform distribution ($\alpha = 1$) is not a good prior, because it leads to weights that are more skewed than the target.
In other words, random fluctuations observed in edge samples are interpreted as actual patterns, which can be viewed as a case of overfitting to the edge observations in this special example.
Although BaCon still performs better than the methods that do not use constraints, in \cref{sec:results:skewed}, we show how we can further improve the inference by taking prior knowledge about the peakness of the distribution into account.

\begin{figure*}[!ht]
     \centering
        \includegraphics{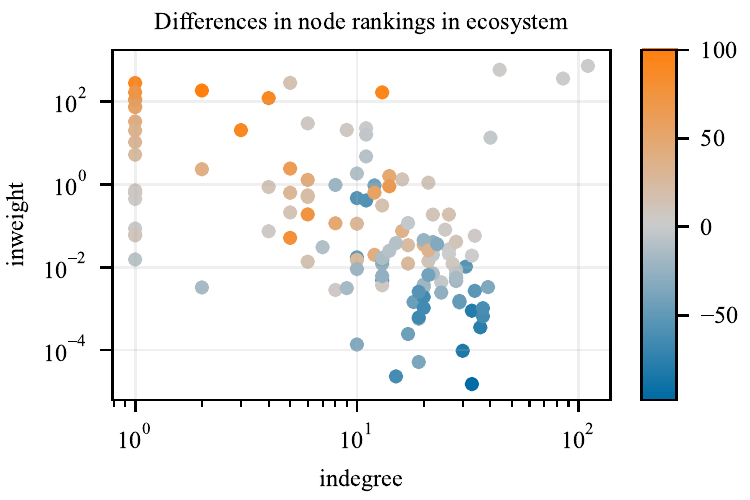}
        \caption{
            Differences in node ranking in the ecosystem network as a function of indegree and inweight. The color encodes the difference between the ranking nodes get based on the ground truth transition matrix $\Tgt$ and the ranking obtained from the unweighted graph topology $G$. Orange nodes are ranked high by the ground truth and low by the unweighted topology and the blue the opposite. Grey nodes have similar rankings with both matrices. 
            The nodes with high (low) indegree but low (high) inweight are ranked opposite in the topological and weight information and are the reason why BaCon under-performed in the \emph{ecosystem} network (see \cref{fig:PageRank_empirical}).
        }
        \label{fig:fig_ecosystem_rankings_explained}
\end{figure*}

The \emph{ecosystem} dataset presents the other extreme.
This network has a highly skewed weight distribution, with ten orders of magnitude between the largest and smallest weights and a Fisher-Pearson coefficient of skewness of $13$~\citep{kokoska2000crc}.
The flat prior weight distribution (cf. \cref{sec:BaCon}) used in the experiments outlined above makes such skewed outcomes very unlikely and leads to weights that are less skewed than the target.
In other words, we are underfitting the observations.
This provides a possible explanation why the frequentist method outperforms BaCon, since the frequentist method exhibits a tendency to overfit the observations.
However, in the \emph{ecosystem} dataset we also observe that the naive Bayesian method leads to better node rankings than BaCon, even though we know from \cref{sec:results:inference} that it underfits the transition probabilities more than BaCon.
The flat prior weight distribution causes underfitting of both BaCon and naive Bayesian methods and thus cannot explain alone the better ranking of the naive Bayesian method.
Therefore, to understand this, we have to consider how the node ranking is impacted by the flat prior weight distribution in conjunction with the topological constraints.
The naive Bayesian method underfits with the prior of a fully connected network, by which all nodes have equal PageRank scores and an undefined ranking.
BaCon underfits with the prior of the unweighted topology; in the topology, nodes have different PageRank scores and their rankings are positively correlated with, but not identical to the ground truth rankings.
Because of the underfitting, the scores of nodes remain close to the ones given by the prior, and therefore BaCon outperforms the naive Bayesian method in the regime of very small data.
However, once we consider a larger number of observed edges, BaCon's prior actually becomes a hindrance.
Since the PageRank scores of nodes in the topological prior are not all equal, they require more data to change sufficiently to improve the incorrect rankings. 
In contrast, the PageRank scores of nodes with the fully connected prior of the naive Bayesian method are all equal and thus need less data to change sufficiently to adjust the undefined ranking to the ground truth one.
We can peek into the mechanism of how inclusion of the topological prior hurts BaCon's node ranking in the \emph{ecosystem} dataset by looking at \cref{fig:fig_ecosystem_rankings_explained}.
In this figure, we plot the indegree and inweight properties of the nodes, and the discrepancies in their rankings.
The high skewness of the weight distribution makes it possible that some nodes have high (low) indegree ($x$-axis) in the topology but low (high) inweight ($y$-axis) in the ground truth weighted graph.
It is exactly those nodes that have the highest differences between their ground truth rank and their rank from the topology (the difference is encoded in the color of the scatter plot markers).
This situation cannot occur with the naive Bayesian because all nodes have the same indegree in the prior (the fully connected graph). 
In \cref{sec:results:skewed}, we will show how to incorporate prior knowledge of the skeweness of the weight distribution in the inference step.

Finally, we discuss threads to validity.
When the information of constraints $G$ is partial and biased, the inclusion of constraints in the inference can negatively impact the downstream tasks.
As an example, take a weighted network in which a node $v$ has the highest centrality and is not connected to all other nodes.
If we only know the constraints on the node $v$, then the resulting (partially known) topological constraint will be a fully connected network except for the connections to and from node $v$. 
Thus, looking only at this topology, we would assume that the node $v$ is the \emph{least} important node in the network, and, since BaCon mixes the information present in constraints and in the observations, it would negatively impact node ranking.
A similar situation can occur with cluster detection in a weighted network with ground truth clusters $C$.
If the knowledge of the topology of the constraint is biased in such a way that there is a different node grouping $C'$ and we only know constraints across clusters $C'$, then the inferred clusters could also be biased towards $C'$. 
In \cref{sec:results:skewed}, we will discuss how our approach allows a practitioner to incorporate the information about biases of the two datasources in the inference.

\section{Tuning the A Priori Distribution of Edge Weights}
\label{sec:results:skewed}

In the previous section, BaCon's results on \emph{ecosystem} showed a subpar performance for node ranking and clustering.
Additionally, on \emph{powergrid}, we observed overfitting in the medium samples regime.
Both observations relate to the characteristics of the datasets' edge weight distributions: highly skewed in \emph{ecosystem} and heavily peaked for {powergrid}.
In fact, prior knowledge on those systems may lead us to expect that the edge weight distributions exhibit such characteristics:
Ecosystems heavily rely on simpler species to introduce energy into the system \citep{chapin2002principles}.
In a power grid, possible voltages are determined by engineering standards.
Leveraging the Bayesian approach of BaCon, we can use these expectations to formulate non-uniform priors for the edge weight distribution. 
As discussed in \cref{sec:BaCon}, by changing $\alpha$ we obtain prior distributions with different shapes: $\alpha < 1$ gives a higher probability to skewed distributions while $\alpha > 1$ assigns higher probability to peaked distributions.
By matching the parameter $\alpha$ to the expected weight distribution in a given system, we can thus include prior information on the edge weight distribution into our inference, which can make it more data-efficient.
In this section, we demonstrate the benefit of this procedure in the \emph{ecosystem} and \emph{powergrid} data sets.

\begin{figure*}[!ht]
 \centering
    \includegraphics{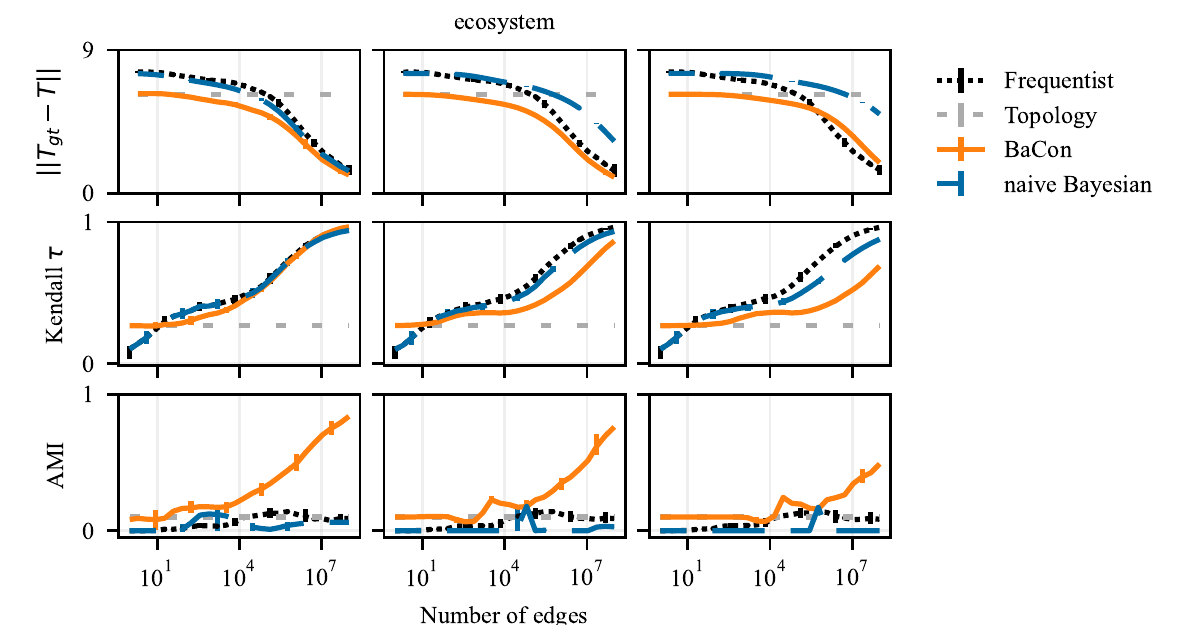}
    \caption{
    Effect of using different concentration parameters $\alpha$ alongside the topological prior on the \emph{ecosystem} network. 
    Rows indicate the task (top to bottom: transition matrix inference, node ranking, clustering), columns the different values for $\alpha$ (left to right: $\alpha = 0.01$, $\alpha = 1$, $\alpha = 10$). We observe that compared to the value used in the previous experiment ($\alpha = 1$) BaCon performance on the \emph{ecosystem} network improves for the smaller $\alpha$ (left column).
    }
    \label{fig:varyingAlphaEco}
\end{figure*}

\begin{figure*}[!ht]
 \centering
    \includegraphics{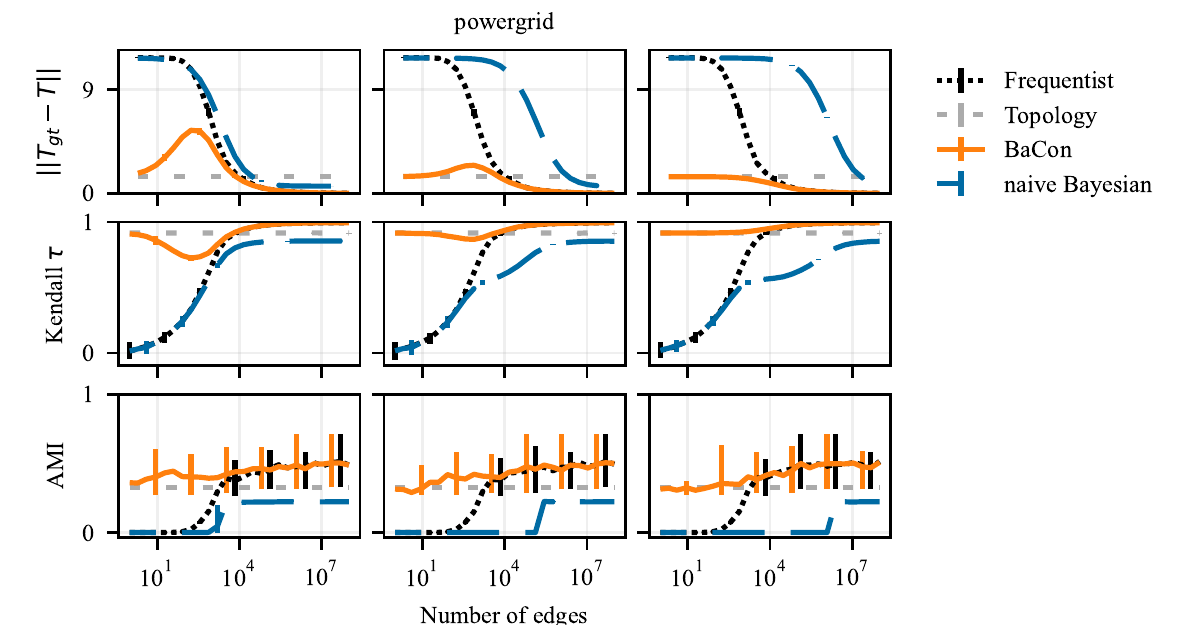}

    \caption{
    Effect of using different concentration parameters $\alpha$ alongside the topological prior on the \emph{powergrid} network.
    Rows indicate the task (top to bottom: transition matrix inference, node ranking, clustering), columns the different values for $\alpha$ (left to right: $\alpha = 0.01$, $\alpha = 1$, $\alpha = 10$). We observe that compared to the value used in the previous experiment ($\alpha = 1$) BaCon performance on the \emph{powergrid} network improves for the larger $\alpha$ (right column).
    }
    \label{fig:varyingAlphaPow}
\end{figure*}

We show the results of the experiments on \emph{ecosystem} in \cref{fig:varyingAlphaEco} and of the experiments on \emph{powergrid} in \cref{fig:varyingAlphaPow}.
The experiments are the same as the ones in the previous section but, in addition to the uniform prior ($\alpha = 1$), we now also consider a skewed ($\alpha = 0.01$) and a peaked prior ($\alpha = 10$).
The performance of our method for three different tasks is shown in the three rows of \cref{fig:varyingAlphaEco} (top row: inference of transition matrix, middle row: node ranking, bottom row: cluster detection).
The columns show the performance for different values of the parameter $\alpha$ (left column: $\alpha=0.01$, middle column: $\alpha=1$, right column: $\alpha=10$).
As expected, \emph{ecosystem} performs best with the skewed prior and worst with the peaked one.
Conversely, \emph{powergrid} performs best with the peaked prior and worst with the skewed one.

\paragraph{Discussion}
Our results show that using BaCon with suitable informative priors eliminates both the poor performance on \emph{ecosystem} and the overfitting for medium-size samples in \emph{powergrid}.
For simplicity, we performed our experiments using two specific values of $\alpha$ ($\alpha=0.01$, $\alpha=10$).
In practical cases, the exact choice of $\alpha$ has to be determined case-by-case and is an open problem for further research.
\citet{nemenman2001entropy} remark that the choice of the concentration parameter $\alpha$ of the Dirichlet distribution which maximizes the variance of the entropy of the multinomial distributions drawn from it is the Schurmann–Grassberger estimator ($\alpha = 1/K$ where $K$ is the alphabet size of the multinomial distribution), albeit they also note that the resulting variance is narrow.
Since the skewness and entropy of a multinomial distribution are related, we suspect that the Schurmann-Grassberger estimator might be the optimal choice of a fixed $\alpha$ that makes the least assumptions about the skeweness of the distribution.
Finally, we highlight that reducing $\alpha$ would help in the case when the constraint is partially known (and possibly biased), considered at the end of \cref{sec:results:analysis}: since spurious edges have zero weights, they effectively increase the skewness of the weight distribution, and thus require lower $\alpha$. 
In this case, the reduction of $\alpha$ could also be interpreted as assigning smaller importance to the topological prior when it is less reliable.
In summary, although the exact choice of $\alpha$ for skewed distributions remains an open question, our preliminary investigation demonstrates that BaCon can incorporate prior information on the edge weight distribution, and that doing so can further improve the inference.

\section{Related Work}
\label{sec:relatedWork}

In our work, we propose a Bayesian method to infer transition matrices from data capturing a possibly incomplete set of repeated interactions between nodes in a graph that are subject to topological constraints.
A related approach has recently been proposed to address the detection of the optimal order of higher-order graphical models for causal paths in temporal networks~\citep{petrovic2022learning}.
While this approach is suitable to address the important problem of model selection in higher-order network models~\citep{lambiotte2019networks}, the challenge of inference and model selection in first-order weighted graphs, which are abundant in practical network analysis, is largely been overlooked.
To this end, our work can be seen as a generalization of the approach in \citep{petrovic2022learning} to standard weighted graphs.
Extending and generalizing this recent work, we further explicitly evaluate the accuracy of the inference in incomplete data and assess its impact on downstream network analysis tasks for several synthetic and empirical data sets. 

Clearly, the inference of transition matrices from repeated interactions is only one particular challenge that fits into a larger body of works addressing other types of network inference problems, i.e. the inference of weighted and unweighted graphs in noisy or incomplete data, or in data with spurious interactions.
A number of works have implicitly addressed the inference of weighted graphs by adapting downstream network analysis tasks to incomplete data, or to data that are subject to errors.
In \citep{smiljanic2020mapping} a Bayesian prior is used to avoid overfitting when detecting clusters in data that are subject to both missing or spurious interactions.
This prior regularizes the community detection results but, differently from the one considered in this paper, does not  use a known network topology as a constraint. 
Similarly, the issue of network inference has been implicitly addressed with the framework of Graph Neural Networks (GNN).
In these works, the inferred network structure and a downstream learning task are jointly optimized by the GNN ~\citep{wang2021graph,franceschi2019learning,jin2020graph,zhang2019bayesian}.
Because of this joint optimization, these works do not aim at a principled inference of an optimal weighted graph model given observations, but rather at finding a network model that optimizes the specific learning task at hand.

Another body of works focuses on inferring the \emph{unweighted topology} of networks from incomplete or biased observations of interactions.
Early works tackle the problem from the perspective of a specific domain like, e.g., social systems \citep{butts2003network} or connectomes in neuroscience \citep{priebe2015Statistical} (further references can be found in \citep{young2020bayesian}).
More recent works define the problem in more general terms as the inference of both topology and a \emph{data model} connecting the topology to the observations.
Here, a \emph{data model} is defined as a model that produces observations as a function of the network topology and additional parameters.
In \citep{newman2018network}, the network topology is obtained through a procedure of expectation-maximization from potentially incomplete or noisy data.
In \citep{casiraghi2017relational}, the problem is tackled from the perspective of the statistical significance of the interactions. 
The method provides a way to infer statistically significant edges from edges that could have happened at random in the system.

Methods similar to the ones above, but using a Bayesian approach for the inference of weighted graphs, were proposed where the posterior is obtained through a Markov Chain Monte Carlo procedure \citep{young2020bayesian,peixoto2018reconstructing}.
In \citep{peixoto2019network}, the prior is a generative model that couples the network inference with the detection of communities. 
In \citep{rabbat2008network} the directedness of the network is inferred from undirected co-occurrence data. 
The data are assumed to be generated by random walks, and Expectation Maximization is used to estimate transition probabilities. 

Unlike the methods above, we assume that we have (at least partial) knowledge on which interactions are possible. 
We further consider situations where no spurious interactions can be observed, i.e. we have access to data that represent a subset of the possible interactions.
We use this knowledge to define a Bayesian prior that, differently from other Bayesian priors in the literature, explicitly constrains which interactions are possible.
We obtain a weighted graph inference method that can be solved analytically, thus being computationally efficient.
Different from existing methods, we also show how the parameters controlling our prior can be tuned depending on the expected skewness of the underlying weight distribution, thus further improving the data efficiency of our method.

\section{Conclusion}
\label{sec:conclusion}

In this work, we address the problem of inferring transition matrices from data capturing an incomplete sample of repeated interactions in networked systems with a known topology.
Examples include data on users navigating information networks with a known hyperlink structure, passengers travelling in a transportation network with known physical topology, or observed social interactions in systems with known social structure.
To the best of our knowledge, no existing inference method has specifically addressed this problem, despite its large practical relevance for network analysis tasks that rely on transition matrices constructed from weighted graphs.

We address this issue with a Bayesian approach, where the prior distribution captures our knowledge of the network topology. 
An experimental evaluation of our method in synthetic and empirical datasets shows that it considerably outperforms a common frequentist inference method and a naive Bayesian approach both in terms of the accuracy of the inferred transition matrix, and in terms of the results of downstream network analysis tasks. 
The prior based on topological constraints regularizes the inferred probabilities and thus prevents overfitting in small data sets (exhibited by the frequentist method). 
It simultaneously limits the degrees of freedom, which prevents underfitting (exhibited by the naive Bayesian approach).
Our results show that such a prior is effective even when the knowledge of the constraints is partial.
Moreover, thanks to its analytical tractability, our approach does not require expensive simulations.
It just requires Bayesian updating based on a simple counting of interaction occurrences.
Highlighting issues in networks with skewed or peaked distributions of edge weights, we finally show how the adjustment of a parameter that controls the prior distribution of edge weights can be used to further improve the data efficiency of our method.

In summary, we propose an intuitive and elegant method for a common problem in network analysis.
Referring to the adjustment of the prior to the expected edge probability distribution, in future works we seek to address the question how a suitable choice of this parameter can be learned from the data. 
We further expect that our method can be used to improve network inference in situations where we do \emph{not} have access to a network constraint, e.g. by combining it with existing methods to infer the \emph{unweighted topology}.
Finally, considering the fact that our method is particularly useful in situations in which the amount of data is small compared to the dimension of the space of possible interactions (i.e. the degrees of freedom determined by the network topology), we expect it to be of considerable interest for the growing community addressing inference tasks in higher-order network models \citep{lambiotte2019networks}.

\section*{Availability of data and materials}
The the code and the results of the current study are available in the Zenodo repository \citep{zenodo_BaCon}.
All real world data sets used in this manuscript are publicly available \citep{veraszto2020whole,ulanowicz2005network,west2012human,FLdata,kim2018depth,peixoto2020netzschleuder}.

\section*{Competing interests}
  The authors declare that they have no competing interests.

\section*{Author's contributions}
    Concept and mathematical foundations: LP.
    Software: LP and VP.
    Experiments: LP and VP.
    Draft manuscript: LP and VP.
    Final manuscript: LP, VP and IS.
    Supervision and project management: IS.
    
\section*{Acknowledgements}
Vincenzo Perri,  Luka Petrovi\'c, and Ingo Scholtes acknowledge support by the Swiss National Science Foundation, grant 176938.

\bibliography{refs}

\end{document}